\documentclass[a4]{article}
\usepackage[a4paper]{geometry}

\usepackage{cite}
\usepackage[cmex10]{amsmath}
\usepackage[pdftex]{graphicx}
\usepackage{array}
\usepackage[tight,footnotesize]{subfigure}

\DeclareMathOperator*{\sumint}{%
\mathchoice%
  {\ooalign{$\displaystyle\sum$\cr\hidewidth$\displaystyle\int$\hidewidth\cr}}
  {\ooalign{\raisebox{.14\height}{\scalebox{.7}{$\textstyle\sum$}}\cr\hidewidth$\textstyle\int$\hidewidth\cr}}
  {\ooalign{\raisebox{.2\height}{\scalebox{.6}{$\scriptstyle\sum$}}\cr$\scriptstyle\int$\cr}}
  {\ooalign{\raisebox{.2\height}{\scalebox{.6}{$\scriptstyle\sum$}}\cr$\scriptstyle\int$\cr}}
}

\usepackage{alltt}

\def\impos{{\;\vcenter{\hbox{\rule{5mm}{0.2mm}}} \vcenter{\hbox{\rule{1.5mm}{1.5mm}}} \;}}

\def\lrarrow{\leftrightarrow \kern-8pt \rightarrow}

\def\2{\frac{1}{2}}

\def\Nav{\langle N\rangle_{\overline T}}

\def\beq{\begin{eqnarray}}
\def\eeq{\end{eqnarray}}
\def\2{\frac{1}{2}}

\def\lrarrow{\leftrightarrow \kern-8pt \rightarrow}

\def\frightarrow{\rightarrow \kern-11pt /~~}
\def\reducesto{\simeq \kern -3pt >}

\def\CAND{\;{\bf AND}\;}

\begin{document}
\newcommand{\strust}[1]{\stackrel{\tau:#1}{\longrightarrow}}
\newcommand{\trust}[1]{\stackrel{#1}{{\rm\bf ~Trusts~}}}
\newcommand{\promise}[1]{\xrightarrow{#1}}
\newcommand{\revpromise}[1]{\xleftarrow{#1} }
\newcommand{\assoc}[1]{{\xrightharpoondown{#1}} }
\newcommand{\rassoc}[1]{{\xleftharpoondown{#1}} }
\newcommand{\imposition}[1]{\stackrel{#1}{\impos}}
\newcommand{\scopepromise}[2]{\xrightarrow[#2]{#1}}
\newcommand{\handshake}[1]{\xleftrightarrow{#1} \kern-8pt \xrightarrow{} }
\newcommand{\cpromise}[1]{\stackrel{#1}{\frightarrow}}
\newcommand{\policy}{\stackrel{P}{\equiv}}
\newcommand{\field}[1]{\mathbf{#1}}
\newcommand{\bundle}[1]{\stackrel{#1}{\Longrightarrow}}

\title{Causal evidence for social group sizes from Wikipedia editing data}
\date{\today}
\author{M. Burgess\\
ChiTek-i AS,\\
Oslo, Norway\\
~\\
and \\
~\\
R.I.M. Dunbar\\
Department of Experimental Psychology\\
University of Oxford\\
Radcliffe Quarter\\
Oxford OX2 6GG\\
UK}
\maketitle

\renewcommand{\arraystretch}{1.4}

\begin{abstract}
  Human communities have self-organizing properties in which specific
  Dunbar Numbers may be invoked to explain group attachments.  By analyzing
  Wikipedia editing histories across a wide range of subject pages, we
  show that there is an emergent coherence in the size of transient
  groups formed to edit the content of subject texts, with two peaks
  averaging at around $N=8$ for the size corresponding to maximal
  contention, and at around $N=4$ as a regular team. These values are
  consistent with the observed sizes of conversational groups, as well
  as the hierarchical structuring of Dunbar graphs.  We use the
  Promise Theory model of bipartite trust to derive a scaling law that
  fits the data and may apply to all group size distributions, when
  based on attraction to a seeded group process.  In addition to
  providing further evidence that even spontaneous communities of
  strangers are self-organizing, the results have important
  implications for the governance of the Wikipedia commons and for the
  security of all online social platforms and associations.

\end{abstract}


\section{Introduction}

Human social groups have a hierarchical structure whose layers have
very specific sizes\cite{dunbar6a}. They form a harmonic series at which the
information flow around a social network is optimized
\cite{dunbar13a,dunbar14a}. The layers emerge owing to both the
interaction frequency with which individuals contact each other
\cite{dunbar11a,dunbar12a} and the sense of trust through increasing
familiarity\cite{trustsummary}. The evaluation of cues for community
membership, known as the Seven Pillars of Friendship \cite{dunbar5a},
and the exchange of tokens have previously postulated this
\cite{dunbar10a,burgesstrust}.  The base of the group hierarchy lies
at approximately 5 individuals. The upper limit at which individuals
have been found to take part in a coherent conversation
\cite{dunbar15a,dunbar4a,dunbar8a,dunbar9a,dunbar16a} is 4, which
appears to be determined by the cost and capacity for humans to manage the
mental states or viewpoints of other individuals \cite{dunbar8a}.

Experiments to scrutinize these notions are notoriously difficult to
perform at scale; however, the community structures surrounding
Wikipedia have provided us with a serendipitous opportunity for a direct test of
these claims, since they gather and publish data in a completely open
manner. In the course of studying Wikipedia in relation to matters of
online trust, we stumbled across a signal in the data concerning the
dynamics of groups that we report on here\cite{trustnotes}.

Individuals come to the Wikipedia platform with a variety of
motivations that ultimately result in the editing of Wiki pages.  This
is a very well documented online process performed by large numbers of
individuals, who may or may not be aligned in their intent. It is one
of the few online platforms or `virtual societies' that openly shares
the histories of editing interactions on its pages.  
There are several classes of agent involved in editing.
About 20\% of changes are made by editorial `bots': these perform
administrative functions like labelling, fixing citations and
correcting spelling, etc. Such change bots have different tolerances
and limits than humans. Editors are usually strangers and often appear
to other users as confrontational figures. This allows us to test
whether the self-organizing principles that structure personal social
networks also apply to groups of strangers and, by implication, to
groups formed for the sole purpose of work. Our study seems to show
that the emergence of highly predictable group sizes depends entirely
on the existence of an initial seed event as an attention attractor,
and an accounting of the economics of that attention as part of the phenomenon of trust.

Wikipedia has a wealth of pages spanning a large number of subjects.
These can be viewed as evidence of a wide variety of intent amongst
users.  Averaging results over large numbers of topics would therefore
be expected to erase any biases associated with specific subject
material, leaving only properties that characterize common aspects of
human participants. Such characteristics would likely be innate to
humans, since the collection of contributors is broad and
internationally diverse thanks to the Internet. In other words,
Wikipedia samples contributions from a wide variety of individuals
with different motivations leading to a high level of entropy in the
writing pool. This grants us the study the universal aspects
of group phenomena with statistical credibility.

The study we report here is thus part of a broader study on the
topic of trust. It reveals an emergent coherence in the sizes of transient groups, which coalesce
to edit the content of the Wikipedia pages. The results demonstrate good
agreement with earlier predictions about social groups.  Groups on
Wikipedia come together episodically and average at around $N=8$
persons during an episode. The probability for a new user to join a
group peaks at around $N=4$ and the distribution has a long
exponential tail.  Working from Promise Theory\cite{promisebook}, we
have obtained a simple formula for this probability distribution of
transient group sizes \cite{burgesstrust}.

\section{Data collection and analysis}

We analyze the statistical mechanics of the time series of changes,
recorded by the Wiki platform, as an underlying causal process and use
very general arguments from dimensional analysis to gauge the
dynamical behaviour of users in the limit of a large ambient
population\cite{scale2}.
In the interests of focus, we shall sketch only the most pertinent
details from the full experiment, i.e. those which contribute to the main result of this
paper summarized by the formula in equation
\ref{formula} and the data fit shown in figure \ref{datafit}.  Full
and detailed descriptions of methodology, process, data, as well as computer
code are shared freely online at \cite{wikidetails} and further
discussion about the theory may be found in \cite{burgessdunbar2}.

The principal source of data for user groups comes from studying the
change history logs, which are maintained automatically and
impartially by the platform software, one for each Wiki subject page.
Analysis of these text logs, parsing their ordered timelines is quite
involved and somewhat time consuming. It involves parsing and
extracting data iteratively. The ultimate outcome is a multivariate
time series for each subject page.

In order to cover a broad mixture of users and subjects in the data, a
selection of Wiki page subjects was chosen randomly to form a base
list of subjects; this was then sub-sampled in batches (in the manner
of a Monte Carlo method).  Answers from partial data were assessed for
stability under varying selections; this helped to keep repeated experiment
executions to within reasonable time limits and establish
repeatability.

Initial tests to determine a suitable ensemble size with statistical
stability resulted in a sample set of around 800 subject pages, which
was then sub-sampled in groups of 100. These pages, in turn, involved
interactions with around 200,000 separate online contributor
identities.  The results, while still noisy by a few percent, exhibit
a robust statistical stability (as shown by the error bars in figure
\ref{datafit}). Our final results are based on the full merged set of
data for completeness.
A wide variety of variables could in principle be extracted from
Wikipedia data. We focus on just a few to demonstrate the main
result of the study.  Table 1 shows the quantities used to assess
the dynamics of the user behaviours. 

\begin{table}[ht]
\begin{center}
\begin{tabular}{|c|c|}
\hline
$L$ & The current total length of the Wiki article being edited in characters.\\
\hline
$N_e$ & The number of users associated with an `episode' $e$.\\
\hline
$\Delta t_i$ & The UTC clock time interval at which an edit was reported.\\
\hline
$\Delta t_i = t_i - t_{i-1}$ & The time interval in seconds between edit $i$ and edit $i-i$\\
\hline
$t_i(e_n)$ & A time stamp identified as belonging to an edit from episode $e_n$\\
\hline
$I(N_e)$ & The number of contention markers counted during the episode.\\
\hline
\end{tabular}
\caption{\small Key quantities in the time series analysis.}
\end{center}
\end{table}
Each article editing history amounts to time series of data, time stamped with a UTC clock
value, and annotated with editing measures and human comments. We look for variables that
offer evidence of users interacting with one another punctuated by periods of rest.
From the theory of trust, we expected contentious interactions between users to
be the significant driver of activity, and this indeed was evident qualitatively
in the data. In other words, users do not appear to be attracted to work together
cooperatively, but rather tend to come in dispute of one another.
Such contention between user identities can be identified by evidence of actions such as tit-for-tat
editing, trigger words in the comments and so on.
The contention intensity function may thus be defined as a piecewise integral of discrete signals
from the history text and difference data:
\beq
I_e(t_i) = 
\left\{
\begin{array}{c}
\text{undos of previous edit} \mapsto +1\\
\text{combative key words in comment} \mapsto +1
\end{array}
\right.
\eeq
This is summed over times relating to an episode, starting from $t_i$ and finishing at $t_f$:
\beq
I = \sumint_{t_i(e)}^{t_f(e)} \; I_e(t)\,dt
\eeq
The episodic or bursty nature of editing implies that users have an inconstant work
rate, rather than making steady plodding increments. A key issue is therefore how to
measure a timeline of such variable rate interactions. Absolute 
work output measures time (as one sees by examining the length of text produced $L$).
Also, time stamps encoded in change logs measure an embedding time. 
Since the discussions are asynchronous dialogues (i.e. not face to
face), one can't infer too much from the precise clock intervals
between responses. A response to one edit could take weeks or minutes.
Rather we look for indications of continuity or discontinuity of
process. Every process thus has its own clock; to assess a process `proper
time', we used the intra-edit interval $\Delta t_i$ and its running
average to look for significant pauses, defined as an absence of
editing over order of magnitude greater than the running average, and
this was then used to identify a scale for episodic cohesion.  

A local punctuation scale was introduced to dissect and splice the timeline
into episodes, in equation (\ref{punc}). If the intra-edit
interval was suddenly much greater than a normal response time, $\Delta t_i \gg
\langle\langle\Delta \rangle\rangle$, i.e. by an order of magnitude, then we consider
that one episode was ended and another was begun: $e_n \mapsto e_{n+1}$
Episodic time was thus identified by the separation of scales measured in UTC time:
where,
\beq
t_i(e_n) \gg t_f(e_{n-1}),
\eeq
i.e. by a punctuation interval which is large compared to the interval between interactions:
\beq
\Delta t(e_n,e_{n-1}) = t_i(e_n) - t_f(e_{n-1}) > \Lambda \,\langle\langle\Delta t_i\rangle\rangle\label{punc}.
\eeq
This is what one would expect in the physics of work: the work rate or `energy' of attention (which is
associated with kinetic trust in the Promise Theory) should be
associated with the timescale of attention processes.
A dimensionless order of magnitude scale factor $\Lambda \sim 10$,
used as a multiplier of the expected interval between edits (denoted $\langle\langle\Delta t_i\rangle\rangle$),
defines for us a notable cessation of activity. In addition, a minimum $\Delta
t(e_n,e_{n-1})$ of one day between episodes is enforced to remove
artificial bursts faster than a normal human attention span.
To adapt to fluctuating editing rates (user edits are not undertaken on a
regular clock basis), a simple convex running average was defined by the recurrence relation:
\beq 
\langle\langle \ldots\rangle\rangle ~:~ \langle\langle\Delta t\rangle\rangle_i = \rho\, \langle\langle\Delta t\rangle\rangle_{i-1} +
(1-\rho) \Delta t_i.
\eeq
A memory retention rate $\rho =0.4$ was used to favour
rate changes within a few interactions\cite{burgessDSOM2002}.

Edit episodes were obviously not of equal length, especially on the
strict wall clock time, but each could be measured in the
characteristic timescale of the page process itself.  The latter was
measured by searching each process for bursts of change and intervals
of cessation, leading to a characteristic timescale.
The correct identification of process time is necessary to
scale activity dimensionally in accordance with dimensional analysis and the Buckingham
Pi theorem\cite{scale2}.  This method is a standard way of eliminating
variability in the magnitude of statistical fluctuations.
For the
finite data sets, the episode's clock rates between activity and
inactivity fall into separable bands with an easily computed average
value. The parameter $\beta$ in equation \ref{formula} serves as a
residual `manual knob' with which to adjust the efficiency of this coupling.

Apart from the identification of episodes as a temporal phenomenon, 
our other key variables is the size of groups associated with these meta-events.
The number of users in a time series can be 
defined by an idempotent sum over distinct user identities over an episode:
\beq
N_e = \text{dim}\; \bigcup_{t_i(e)}^{t_f(e)} \; \text{username}(t_i)
\eeq
It's unknown how many different identities real persons might have, but this will not affect
the analysis as long as no user impersonates more than one individual
during a particular editing episode. 
The number of edits per episode is not a relevant measure here, but is represented
indirectly through the contention numbers. Furthermore, there is no measurable role
played by the subjects themselves in biasing the results.
By eliminating the times from these quantities, we can represent the contention as a
function of the group size $I(N_e)$ per episode, and further consider the distributions of these.

Once it became clear that edits involved finite episodic groups of
users, we need to know whether groups were formed by interactions of order $N^2$ or
of order $N$. If group members had no preferred member, then one would
expect a closer fit to a complete clique of order $N^2$. Conversely,
with an interaction seeded or led by an initial individual (a strong
leader or just someone who starts the ball rolling), the process of
group attraction would be a star network of order $N$. Order $N$
variance was markedly less than for plotting against $N^2$, suggesting
that once an episode has begun, the episode itself acts as an
attractor for uncoordinated individuals to arrive and join in. This
was even more strongly evident in the shape of the probability
distribution.  A dependence on $N$ rather than $N^2$ is reflected
directly in equation \ref{formula}, based both on theory and empirical
fit.

There are several checks on can undertake to verify the consistency of these choices, including
measuring the correlation between text length $L$ and the total number of users
counted in the log for each article to see if some users are responsible for
most of the text or not. Although this does not necessarily affect the history of interactions
surrounding the editing, it speaks to the potential biasing of results around specific 
`strong' individuals. Figure 1 shows this relationship to be a
narrowly confined power law across all the articles, as one might expect for a random
variable with high entropy. A fairly constant rate of users is
involved per article length.  The average duration of an episode
expended per number of involved users is plotted in logarithms in
figure 2. There is thus a rough correspondence between the length of a Wiki article and
the total number of users involved in its creation, as one might expect both
to be measures of process work (effort). Conversely, the level of contention
between users was quite independent of subject matter.  Our
measure of contention in the editing process is flagged both by
argumentative trigger words in the log text, as well as direct actions
such as undoing a previous user's efforts combined into a single
counter. 

We point out that the subject matter of Wiki pages was at no time relevant for
the group dynamics.  We were able to infer that---over the total
sample---subject matter acts mainly as a seed for attracting certain users.
Once these users arrived, they tended to behave quite uniformly.
If one delved into the psychology of individual responses one might find
hidden depths to interactions, but on a statistical level behaviour is ruled
by simple signalling of an intent to engage contentiously.
We saw no particular correlation between user behaviour and subject
matter.  There is as much contention for subject pages on cabbages or
mathematics as for those on politics and celebrity adulation, provided
they were sufficiently long lived.  This is presumably because the
reasons for contention and criticism are human-led qualities and thus
somewhat similar across all subject areas. More interesting, though
not surprising, we found that human behaviour differs from machine bot
behaviour in this regard. Excluding machine contributions (which are
easily identifiable by name) brought the probability distribution
closer to the predictions made in earlier work. Including them led to
a small shift of supportable group sizes, as machines can tolerate
contention better than humans.
In summary:
\begin{itemize}
\item We find that a high level of entropy in page characteristics and
  contributors is consistent with sampling invariance and use this to
  adopt a Monte Carlo approach, sampling random groups of pages and
  averaging the results.  Since analyzing the pages is time consuming,
  even for a computer, this allows us to infer a high confidence in
  the consistency of results from independent random samples.  We then
  derive the final results from 827 subject pages and just over
  260,000 involved users.

\item Users come to edit a subject page that they care about more or less
  {\em ad hoc}.  Once one author has arrived, others will notice the
  changes and join in. These begin to form a group to either challenge
  or improve on one another's contributions, depending on the
  alignment of their intentions.

\item Editing takes place in bursts of activity, punctuated by longer
  gaps. We can thus identify a series of ``episodes'' for every page,
  which we take to be causally independent. Each episode yields a
  group size and duration and a level of contention that one sees from
  the signals embedded in the historical record.

\item For every subject page there is also a history page which we can
  read and analyze using appropriate software. The pages show evidence
  of user identities, which might be transient and even deliberately
  anonymous but which are assigned regular identifiers by the
  platform. We can see which edits were contested by others and the
  sizes, dates, and times of the edits. This allows us to construct a
  process timeline and to identify contention between individuals.
  Users can argue and even undo one another's changes, and these
  events can be counted.

\item There is little evidence of coordination between the
  contributors across episodes. Coordination is essentially a
  stigmergic process via the medium of the editing produced. In part,
  it is also a confrontational process, as we see from the history and discussion pages.

\item Contention between users can be identified when one user
  undoes the contributions of another, rather than adding to them.
  From even a cursory inspection of results, it's clear that
  contention between users is a significant issue in editing on all
  pages, so special focus was attached to the level of overlap between
  edits and on cases where one user undid the changes made by another.

\item We also note that, in excess of 20\% of all edits were made by
  automated ``bots'', some of which are designed to patrol changes for
  such activity (as well as trivial issues like spelling corrections),
  so the intentional aspects of behaviour are to some degree invoked
  by proxy.
\end{itemize}

\section{Dynamical factors in Wikipedia editorial clusters}

\begin{figure}[ht]
\begin{center}
\includegraphics[width=12cm]{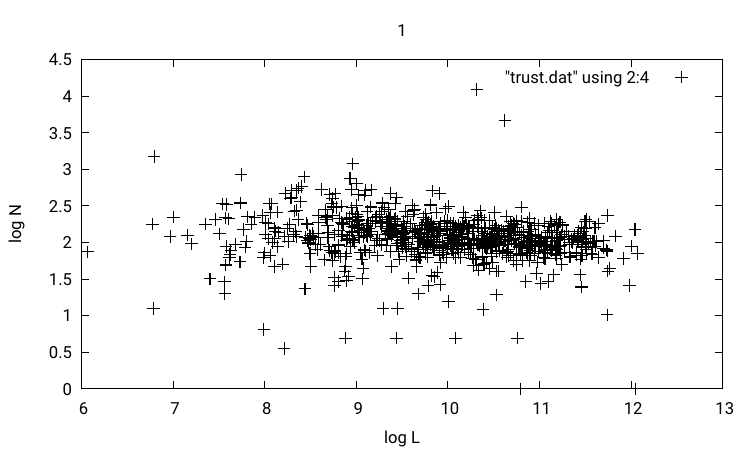}
\caption{\small Relationship between the average number of users
  contributing per episode as a function of article length. $\langle
  N\rangle(L)$ is also noisily constant or decays in a weakly
  exponential manner, suggesting that there is a more or less constant
  stream of new battles to be fought or issues to be solved. The (natural) log
  plots show a weak tendency to exponential behaviour, as expected
  from a probabilistic arrival process. However, the restricted number
  for $N$ evident in all the graphs makes this almost
  insignificant.\label{totgr12}}
\end{center}
\end{figure}

Figure \ref{totgr12} shows how the number of users involved per
editing episode varies with the length of the main
article at the time of snapshot. The article length may be taken as a heuristic 
proxy for the total time
invested by all editors of the page's history.  We see that rate of
user comings and goings decays almost imperceptibly with the elapsed time or length,
suggesting a slow stabilization of subject pages. However, this is small
compared to the level of activity maintained over time. In essence, we
see little evidence of pages ever being finished.

All the pages we examined appear to be subject to continuous
contentious editing. The rate of stabilization with length (time) is
quite slow, as can be seen using the text length of the article as a
proxy for total time elapsed since creation. The bursty nature
evident in the data suggests that causal behaviour is limited to
individual episodes, so we treat each episodic burst as a separate
event.

\begin{figure}[ht]
\begin{center}
\includegraphics[width=12cm]{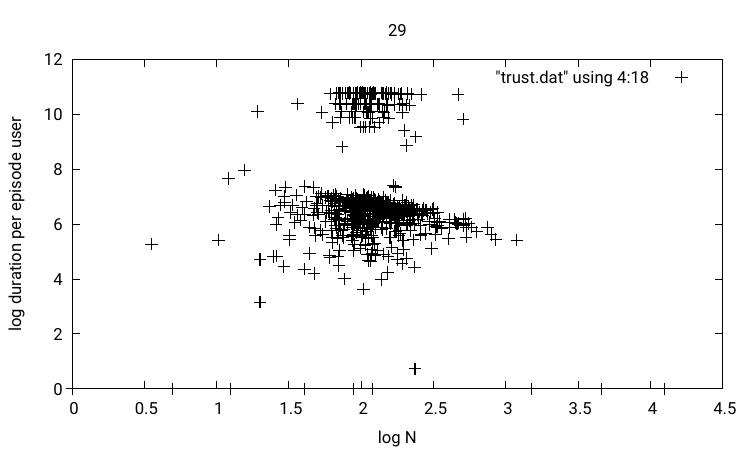}
\caption{\small Duration (lifetime) per episode versus $N$ in rescaled episode time. This
  is the average fractional time spent by a single user over an
  episode. The duration is measured in days. The graph axes (natural
  logs) all show clustering in a region, rather than a line
  relationship, suggesting that the tendency to continue interacting
  (fighting or discussing) decays weakly and exponentially with the
  number of users, but is almost constant. The number of users itself
  does not vary much in episodes, so the cluster is
  narrow.\label{totgr272829}}
\end{center}
\end{figure}

The lifetime over which a group edits is closely bounded
in duration, although the bounds are very noisy. Figure
\ref{totgr272829} shows the existence of at least two distinct classes
of episode duration evident in widely separated clusters, but these are not
distinguishable by our counting method. This not an
artefact of the analysis: the same feature recurs in different sample
sets, although most of the activity is in the lower band. In effect,
the size of an active editorial group remains surprisingly constant
despite considerable turnover in membership over time. It remains to
be studied whether there is significant turnover in the groups with
very long durations.  This would tend to make sense, given that the
size of the group remains statistically constant over even
exponentially longer interactions.

From the $N$ dependence of data, we infer that users are attracted to
an editing episode by the initiation of changes made by a user seed.
In other words, some user comes at random to perform work and that
attracts others to see what's going on. The number of users in a
typical episodic event is noticeably clustered. We see this in a few
measures, the most interesting of which is the plot of contentious
changes versus size of users in an episodic cluster. An average group
size associated with contention is around $N=8$ (with a variation
between 7.75-8.2, depending on how we calculate and sample the
average). We take this to be the group size at which contention
reaches its maximum limit. Note, however, that upwards of 20\% of all
users on the platform were approved bots working by proxy on behalf of
both the platform and others. If we exclude bot activity, there is a
small reduction in average group size that brings the data fit closer
to the numbers widely discussed in the literature.

\begin{figure}[ht]
\begin{center}
\includegraphics[width=12cm]{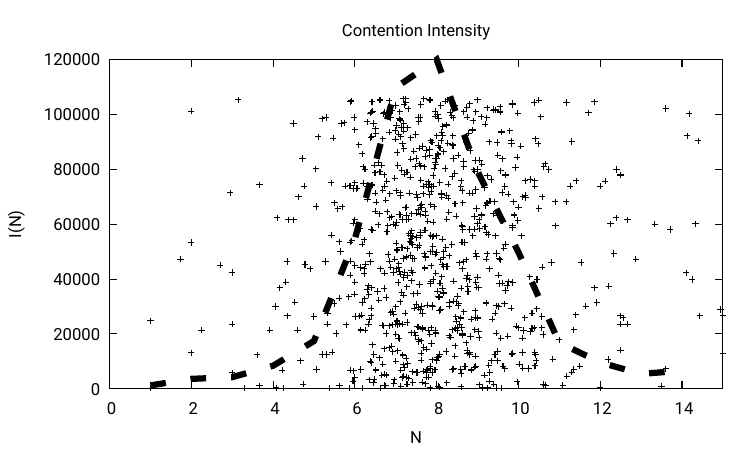}
\caption{\small The contention $I(N)$ or intensity of mistrust, peaks
  around $N=8$ (with a variation between 7.75-8.2). The scaled probability
  density for the raw data is superimposed as 
a dashed line for visual aid.  A more nuanced value for $N$ can be found as
  an expectation value of this distribution to account for the
  uncertainty.  The vertical cutoff in history length accounts for the
  flat top in data point ceiling.\label{zoom2}}
\end{center}
\end{figure}

Figure \ref{zoom2} shows an enlargement of the non-empty region of the
plot of contentions per event as a function of users per episode.
This shows that, although the rate of contention varies somewhat
between subject pages, the size of the group and the level of
contention is confined to a relatively predictable region of parameter
space in the range $N=[0,15]$, centred on $N=8$. There are no other
clusters evident.  This tells us both that contention is intrinsic to
users and that users form groups of predictable size.  This suggests
that there is single causal explanation for this cluster size.  

\begin{figure}[ht]
\begin{center}
\includegraphics[width=12cm]{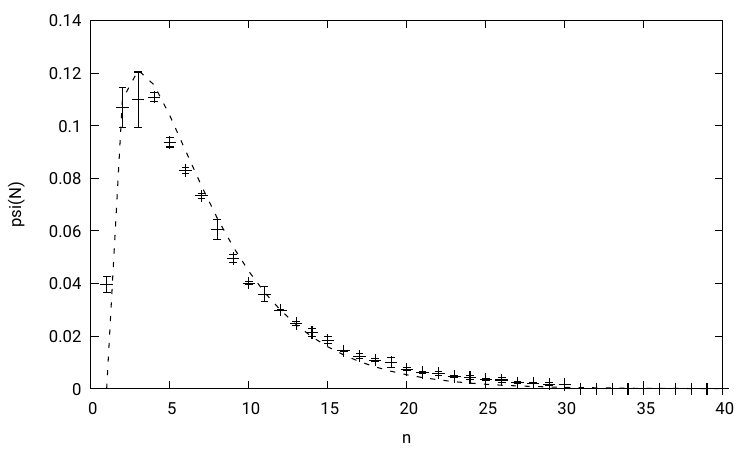}
\caption{\small The principal result: a normalized frequency spectrum of group sizes
  indicates the probability of finding an episodic group of size $N$.
  We note that the peak near $N=4$ is well below the group size that
  maximizes contention at $N=8$.\ref{formula}. The stippled line shows
  a theoretical fit for the curve derived from Promise Theory and
  dimensional considerations (see text below). Crosses indicate
  estimated error uncertainty. Even small changes to the functional form of
this curve lead to significant deviations from the data, which is suggestive that
the theory is basically sound.\label{datafit}}
\end{center}
\end{figure}

The most striking evidence of group size coherence may be seen in the
frequency plot, however. In figure \ref{datafit}, crosses show actual
data and the dotted line a theoretical fit. The unexpectedly close fit
suggests an underlying determinism, albeit on a statistical level. The
spectrum of group sizes across all episodes is remarkably free of the
noisy artefacts from specific measurements, so we expect it to be
quite robust. Error bars are no larger than the crosses.  It has a
classic decaying exponential tail typical of entropic processes.  More
significantly, it also has a non-exponential growth feature for small
$N$, which is where we expect the interesting dynamics to be.

Taking different sample sets at different times and of different sizes
had a negligible effect on the values for the averages used in the
scaling relation, as one would expect for universality relations.
Clearly the noise levels could vary. However, since we rescale by
average noise scale in order to compare unequal processes, this is largely
eliminated. Such rescaling is typical of how one eliminates noise in statistical
mechanical formulae out of equilibrium.

The implications of this for group sizes is that universal
characteristics do not depend on the details of each process but
rather on the core rates of interaction relative to the agents
involved. Only by eliminating time as a variable can we hope to
compare the innate properties of widely different agents according to
their core commonalities. We speculate that those would be the
cognitive or interaction process invariants, and the order of
magnitudes of the ratios seem to confirm that both represent typical
sociological and neurological values.

Note that our analysis was not trying to optimize a regression fit of
pre-classified data onto an off-the-shelf distribution, as one
might hope to do for a known process like a questionnaire. Human
conversations, after all, are not processes based on completely random
variables.  Given what we couldn't know, we have kept to somewhat
conservative methods, principally using dimensional analysis to look
for scale independent effects, which might be compared to
theory\cite{burgessdunbar2}.  There is no other reason to expect a
cleanly functional form for the process of episodic interaction.

\section{A model of group formation based on mistrust}

We were able to explain the results in terms of a model that combines the roles
of user intent and the work effort involved in coming together. To do
this, we measure the amount of semantic ``novelty'' or ``attention
provoking content'' in both the Wikipedia article texts and the
supporting editing histories. We define the amount of work done by
users according to the length of their text changes in alphabetic
characters (for glyphs we can also count the numbers of strokes),
since short or easily written words are generally more common.

Since we know that users join episodic groups coincidentally, randomly
searching and only later following if they are sufficiently attracted
to participate (they do not receive notifications or marching orders),
the story suggested by this spectrum is that users accrete into groups
`gradually', based on subject and activity. On reaching a certain
critical size they then fizzle out.  The timescale for this process is
not constant, but we are rescaling the episodes according to the
average rates for each subject, as measured in advance.  

We see that, initially, the probability of adding new users to a group
begins to falter around a maximum of $N=4$. In other words, users are
less inclined to join a discussion that already contains four persons.
By the stage at which contention is maximal ($N=8$), the probability
of new attachments is already significantly reduced. Since we know
(from the timeline data) that the group episodes terminate, we
conclude that the entire group disbands.  Thus, the groups themselves
decay over their proper time durations, probably as a result of
general entropic diffusion. Such exponential decay is a typical
statistical feature of high entropy mixing.

The amount of work expended by users is linked to the amount of
attention offered by each agent to the conversational interaction, not
to the amount of text added. This fact plays a key role in explaining
the results in terms of mentalizing and mistrust. The amount of
activity invested appears stimulated by what we might call mistrust of
the changes made by others rather than the amount of it. We can assume
that few if any of the agents know one another. There is certainly
little evidence of explicitly cooperative behaviour.
Over time, one expects all text might be altered
in major and minor ways, but when this occurs in the manner of a duel
between two users, it marks a significant semantic interaction. In
such a case, we characterize the behaviour as ``contentious''.

Let us assume that there is a reservoir of agents whose intentions are
distributed with high entropy. Wikipedia does not advertise nor
incentivize editing in any way, so the arrival of a single agent to
edit a page is completely ad hoc. Using Promise
Theory\cite{promisebook}, we can assume that a topic is
represented by a direction in the space of promise outcomes and that
there will be a subset of the agents who are aligned with their
understanding of this approximate topic.
The arrival and departure of users now indicates a kind of
`detailed-balance' model for the statistical stability of the group
spectrum, i.e. the probability of finding a group of size $N$. We can
assume, from the earlier remarks, that the set of users affiliated
with the Wikipedia platform is large and broad in its characteristics,
i.e. it has effectively maximum entropy at large $N$. 

Suppose that a new Wikipedia topic page is started at random by any user or agent. Eventually
another agent will find the changes and be attracted by the seed of a
common interest. If users trusted one another there would be no need to look,
but if they don't then they invest time to verify and possibly alter the text.
This hub attraction leads to kinetic work activity proportional
to $N-1$.

To determine the rate of attention for the agents, we follow the
guidance offered by the dimensions of work/energy for work done by a
pressure or force and the resultant kinetic response. Precise details
needn't be known to find the rate of temporal evolution as a velocity
change from the kinetic work ($1/2mv^2$).

Given any level of attractive potential for curious agents, the
rate of kinetic attachment would be expected to flatten like the square root
of that potential investment of work.  From information theory, maximal entropy in a
conservative reservoir of agents generates a Boltzmann distribution $\exp(-\beta
E)$, where $E$ is a work-energy parameter. We don't actually
require the number of possible agents to be conserved, as the number is assumed
to be essentially infinite. A parameter $\beta$ represents the
agents' average intolerance for contention `back pressure'. Choosing the
only dimensionless combination of parameters $\nu \sim \beta(N-1)/\Nav$,
where $N-1$ is the group size exerting pressure on a newcomer and
$\Nav$ is contention maximum groups scale, we obtain a probability of

\beq
Pr(N) = Pr(\text{attachment}) \CAND Pr(\text{dissipation}) \sim \sqrt{\nu} \times \exp(-\beta \nu).
\eeq
From this, we can derive an expression for the probability of
finding a group of size $N$: 
\beq
\psi(\nu) = \frac{4}{\sqrt\pi}\; \frac{\nu^{\2}\; e^{-\nu}} {\langle N\rangle_{\overline T}}, 
~~~~~~\nu = \frac{2\beta(N-1)} {\langle N\rangle_{\overline T}} ~,~ (N > 1).\label{formula}
\eeq
The broad narrative implied by this result is that new pages are
random events (`event one' of an episode).  These events form a seed
that attracts the attentions of others.  In a group of total size $N$,
$N-1$ will be stimulated by the seeding to follow the rising square
root rate of kinetic attention. This corresponds to `grooming' work in
the language of \cite{dunbar1}.  As the back-pressure arising from
inevitable contention in the group rises, innate limitations on
agents' capacity for this work drives them away. The entropy of the
total mixture ensures this is exponential on a large scale.

The square root growth term is a consequence of the work/energy invested in
attending to the ongoing process. The cognitive cost of increased
familiarity is an increased processing time cost \cite{dunbar3a}. In
physics, energy is the complementary variable (and thus generator) of
temporal evolution in a system. This relationship is essentially
information theoretic on a statistical level. In this respect,
information is a cybernetic concept in the sense that it can involve
the exchange of actual information about some aspect of the world or
some token such as grooming that defines the quality of a relationship
(see also West et al. \cite{dunbar13a,dunbar14a}). The link between
intentions, promises, and trust points to this intentionality as the
seed attractor in the explanation of process network dynamics
\cite{trustnotes}. The attractive `force' during growth is not a
network effect but a kinetic attention effect over these networked
bonds.  Attraction does not therefore necessarily imply proximity in
physical spacetime, but rather in intention space (i.e. the alignment
is in intention first and in position only as a secondary consequence
of attention \cite{trustnotes}).

We need to be careful in interpreting statistical distribution laws as
causative, because a statistical law does not necessarily have reverse
causal implications for individual instances. A tolerance for group
size derived statistically may only be a weak indicator of individual
behaviour. However, there is something intriguing about this result on
a number of levels. There is an `invisible hand' style shaping of
outcomes which is not quite a force, but which could be modelled as
one in an effective theory. Indeed, theory suggests that there is a
relationship between these outcomes and innate properties.

The distribution in figure \ref{datafit} relates the tendency for contention in a group to a
tendency for neighbours to tolerate one another's presence. Initially,
this work seems to be the very essence of the process of discussing
and contending over subject matter editing changes. Later, this
progress seems to be overpowered by the tendency for the costs of
contention to overcome the perceived benefit.  This interpretation
makes sense because the contention maximum always lies at larger size
than the most common maximum size. In other words, contention is still
growing when the group starts to falter.

\section{Universality of the distribution and bots}
It is tantalizing to speculate on the scaling
implications and the wider behaviour of the curve in figure
\ref{datafit}. Implicit in its functional form is a kind of stepping
stone hierarchy of scales, determined in detail by the contention cost
$\beta$.  
Although we have only obtained data here for $\Nav=8$, but we can
nonetheless extrapolate the consequences of the model for other values
with data from other sources.  If we examine the relationship between
the maximum of contention and the maximum equilibrium group size, we
see an interesting hierarchy of sizes, reminiscent of the Dunbar
hierarchy (5,15,...150,500, etc.).  The precise values depend on the
choice of $\beta$, whose value lies close to 1 for this work, but
tolerates minor adjustment to smaller values. The value that generates
the usual Dunbar hierarchy of sizes lies between $0.875$ and $0.95$.

Implicit in the statistical law (\ref{formula}) is a prediction that
there is a fission rate for groups with contentious interactions of
approximately $2\beta$ for $N \gg 2 \beta$ (for smaller $N$ the
integral nature of $N$ precludes a precise analytical expression). The
values in this study are consistent with the work on conversations
summarized in \cite{dunbar3}. These show conversational groups of
around $N=8$ rapidly partitioning into two separate semi-independent
subgroups.  The emergence of a contention limit or peak level plays
the role of an `innate' characteristic for the bulk of the agents. We
should not forget that 20\% or more of agents are in fact software
bots, with potentially infinite tolerance, whose behaviour is largely
triggered by human activity.  The role of attractors is a key feature,
represented by the factors $(N-1)$ in the spectrum.

In the case of the $\beta=0.875$, we have a series close to
5,15,...150, etc. A group that could loosely contend at $n=550$ would
tend to break up into fragments of $150$, and a group of around 15
might break up into smaller groups of order 5 under pressure, and so
forth. This suggests an interesting kinetic link between the dominant
group scales. Again, we emphasize that these are statistical
tendencies not causal rules at the network level. The network
causality implicit in the formula is one based on the kinetic work
done by the process supported by the local network. The precise
numbers are inevitably subject to a degree of statistical uncertainty,
but the essence of the story is compatible with the scaling hierarchy
of Dunbar numbers known from wider sources. Removing all bot
interactions from the data alters average $N$ slightly to give an
effective value of $\beta=0.93$, which is closer to this idealized
series.

\section{Discussion}

The signature of an apparently universal group dynamics is evident
within the data mined from Wikipedia.  Discovering such a beautiful
statistical curve, as seen in equation (\ref{formula}) in a human process,
is a rare opportunity. The difficulties of meaningful reproducibility
faced by the social sciences make it difficult to find innate
phenomena that can rival those of physics or chemistry. Yet, here we
find a remarkably robust expression of scaling.  We did not set out to
look for it, yet it seems to dominate the behaviour of contributors
and offers a fortuitous opportunity to measure details normally not
available to onlookers. 

We should be clear that, while our theory might have wider
implications, the data we analyze here only provide evidence for a
single phenomenon, with characteristic scales $n_\text{max}\simeq 4$
and $\Nav \simeq 8$.  Nonetheless, the scaling treatment suggests a
universality that might extend to other groups too.  Natural
conversations have an upper limit at four members
\cite{dunbar15a,dunbar4a,dunbar8a,dunbar9a,dunbar16a}, with this limit
seemingly set by the mentalizing capacities of the average adult
\cite{dunbar8a}. Other processes that follow the same {\em dynamical
  similitude} could be viewed as analogous. We discuss this
elsewhere\cite{burgessdunbar2}.

The question we are left with is what is the seeding force or
`invisible hand' shaping attraction and repulsion in the group
processes. Is it the charismatic leader, the threat of predation
(animal or human), the need for cooperation, etc. The formula derived
from Promise Theory suggests there will be one, eventually represented
by the proxy of the group itself.  For Wikipedia, it is clearly the
suspicion that someone is changing a subject others care enough about
to defend. In this sense, the process of `grooming' or maintaining the
relationship is with the subject matter rather than the others in the
group directly, and it is clearly a kinetic one: the activity is that
of a standalone agent, which is therefore limited by the
characteristics of the agent, some of which are innate and others a
function of state.

One of the motivations for studying user behaviour in informatics is
to understand the security of online platforms such as Wikipedia.
Such platforms are increasingly prevalent features of our lives. In
Informatics, trust is usually handwaved away with very simple
questions about identity reliability (trustworthiness), leading to
misleading calls for `zero trust' behaviour. Trustworthiness is a
trait that an agent or individual may have in some quantity (usually
varying probabilistically between 0 and 1) that characterizes you as
an individual \cite{Barrio_2015}. Trustworthiness (usually recognized
by cues of various kinds) can act as a `first pass' basis for
action, but it is at best a broad-brush cue based on generalities.
Trust, on the other hand is not a trait but a process that is built up
over time through repeated interactions: I trust you because I have
interacted with you on a number of occasions and found you to be
reliable (as an individual). The challenge in using trust as a
characteristic in the security of systems is that it represents
something taking shape over very large numbers of events. This may
well make it useless as a practical tool unless there are truly
universal cues that transfer from one case to another.

There has been a tendency to look to `complexity science' to explain
phenomena that relate to biological and social systems. We see little evidence
of that in the results here. At the scale of bulk measurement, we find
remarkably linear stability. This is an indication of a separation of
timescales between interior processes involved in relationship
maintenance and the exterior `boundary conditions' of the social
network.  What is challenging for monitoring systems in general is
that the timescales over which significant changes may occur must be
faster than the timescale over which statistical learning takes place.
This is surely why attentiveness is so expensive and group sizes are
constrained in a hierarchy of tradeoffs.  Indeed, if we seek to
compare singular interactions to an average profile of bulk behaviour,
we are faced with a learning challenge if we are to weed out bad
actors. This challenge is precisely the reason why brains evolved
their phenomenal capabilities in the context of ever larger social
groups. Not surprisingly, it is currently under scrutiny in connection
with Machine Learning for Artificial Intelligence (AI).

\bigskip

Acknowledgment: MB is grateful to Gy\"orgy Busz\'aki for discussions about the neuroscience.
MB's work was supported by the NLnet Foundation and 
RD's research was funded by an ERC Advanced Research grant (\#295663).

\bibliographystyle{unsrt}
\bibliography{bib,spacetime}

\begin{thebibliography}{10}

\bibitem{dunbar6a}
R.I.M. Dunbar.
\newblock Structure and function in human and primate social networks:
  Implications for diffusion, network stability and health.
\newblock {\em Proc. R. Soc. London}, 476A:20200446, 2020.

\bibitem{dunbar13a}
B.J. West, G.F. Massari, G.~Culbreth, R.~Failla, M.~Bologna, R.I.M. Dunbar, and
  P.~Grigolini.
\newblock Relating size and functionality in human social networks through
  complexity.
\newblock {\em Proceedings of the National Academy of Sciences},
  117(31):18355--18358, 2020.

\bibitem{dunbar14a}
B.J. West, G.~Culbreth, R.I.M. Dunbar, and P.~Grigolini.
\newblock Fractal structure of human and primate social networks optimizes
  information flow, 2023.

\bibitem{dunbar11a}
I.~Tamarit, J.A. Cuesta, R.I.M. Dunbar, and A.~S{\'a}nchez.
\newblock Cognitive resource allocation determines the organization of personal
  networks.
\newblock {\em Proceedings of the National Academy of Sciences},
  115(33):8316--8321, 2018.

\bibitem{dunbar12a}
I.~Tamarit, A.~S{\'a}nchez, and J.A. Cuesta.
\newblock Beyond dunbar circles: a continuous description of social
  relationships and resource allocation.
\newblock {\em Scientific Reports}, 12:1--11, 2022.

\bibitem{trustsummary}
M.~Burgess.
\newblock Trust and trustability: An idealized operational theory of economic
  attentiveness.
\newblock {\em preprint paper (DOI: 10.13140/RG.2.2.26862.28480/1)}, April
  2023.

\bibitem{dunbar5a}
R.I.M. Dunbar.
\newblock The anatomy of friendship.
\newblock {\em Trends in Cognitive Sciences}, 22(1):32--51, 2018.

\bibitem{dunbar10a}
A.J. Sutcliffe, R.I.M. Dunbar, J.~Binder, and H.~Arrow.
\newblock Relationships and the social brain: integrating psychological and
  evolutionary perspectives.
\newblock {\em Brit. J. Psychol.}, 103:149--168, 2012.

\bibitem{burgesstrust}
J.A. Bergstra and M.~Burgess.
\newblock Local and global trust based on the concept of promises.
\newblock Technical report, arXiv.org/abs/0912.4637 [cs.MA], 2006.

\bibitem{dunbar15a}
R.I.M. Dunbar, A.~Marriott, and N.~Duncan.
\newblock Human conversational behavior.
\newblock {\em Human {N}ature}, 8:231--246, 09 1997.

\bibitem{dunbar4a}
G.~Dezecache and R.I.M. Dunbar.
\newblock Sharing the joke: The size of natural laughter groups.
\newblock {\em Evolution and Human Behavior}, 33:775--779, 11 2012.

\bibitem{dunbar8a}
J.A. Krems, R.I.M. Dunbar, and S.L. Neuberg.
\newblock Something to talk about: are conversation sizes constrained by mental
  modeling abilities?
\newblock {\em Evolution and Human Behavior}, 37(6):423--428, 2016.

\bibitem{dunbar9a}
C.~Robertson, B.~Tarr, M.~Kempnich, and R.I.M. Dunbar.
\newblock Rapid partner switching may facilitate increased broadcast group size
  in dance compared with conversation groups.
\newblock {\em Ethology}, 123:736--747, 2017.

\bibitem{dunbar16a}
M.~Dahmardeh and R.I.M. Dunbar.
\newblock What shall we talk about in farsi?: Content of everyday conversations
  in iran.
\newblock {\em Human Nature}, 28(4):423--433, 2017.

\bibitem{trustnotes}
M.~Burgess.
\newblock Notes on trust as a causal basis for social science.
\newblock {\em SSRN Archive, available at
  http://dx.doi.org/10.2139/ssrn.4252501 (DOI: 10.2139/ssrn.4252501)}, August
  2022.

\bibitem{promisebook}
J.A. Bergstra and M.~Burgess.
\newblock {\em Promise Theory: Principles and Applications (second edition)}.
\newblock $\chi tAxis$ Press, 2014,2019.

\bibitem{wikidetails}
M.~Burgess.
\newblock Github trustability repository and experimental materials.
\newblock https://github.com/markburgess/Trustability.

\bibitem{burgessdunbar2}
M.~Burgess and R.I.M. Dunbar.
\newblock A promise theory perspective on the role of intent in group dynamics.
\newblock {\em arXiv:2402.00598 [cs.SI]}, 2023.

\bibitem{scale2}
G.I. Barenblatt.
\newblock {\em Scaling}.
\newblock Cambridge, 2003.

\bibitem{burgessDSOM2002}
M.~Burgess.
\newblock Two dimensional time-series for anomaly detection and regulation in
  adaptive systems.
\newblock {\em Lecture Notes in Computer Science, IFIP/IEEE 13th International
  Workshop on Distributed Systems: Operations and Management (DSOM 2002)},
  2506:169, 2002.

\bibitem{dunbar1}
R.~Dunbar.
\newblock {\em Grooming, Gossip, and the Evolution of Language}.
\newblock Faber and Faber, London, 1996.

\bibitem{dunbar3a}
T.~D{\'a}vid-Barrett and R.I.M. Dunbar.
\newblock Processing power limits social group size: computational evidence for
  the cognitive costs of sociality.
\newblock {\em Proceedings of the Royal Society B: Biological Sciences}, 280,
  2013.

\bibitem{dunbar3}
R.~Dunbar.
\newblock {\em Friends: Understanding the Power of our Most Important
  Relationships}.
\newblock Little, Brown, 2022.

\bibitem{Barrio_2015}
R.A. Barrio, T.~Govezensky, R.I.M. Dunbar, G.~I{\~{n}}iguez, and K.~Kaski.
\newblock Dynamics of deceptive interactions in social networks.
\newblock {\em Journal of The Royal Society Interface}, 12(112):20150798, nov
  2015.

\end{thebibliography}

\end{document}